\documentclass[12pt,preprint]{aastex}
\def\b{\begin{equation}}
\def\e{\end{equation}}
\def\l{\left}
\def\r{\right}
\begin{document}
\title{The Size of the Longest Filaments in the Universe}
\shorttitle{The Size of the Longest Filaments}  
\author{Somnath Bharadwaj}
\affil{Department of Physics and Meteorology, Center for Theoretical Studies, Indian Institute of Technology, Kharagpur 721 302, India}
\email{somnath@cts.iitkgp.ernet.in}
\author{Suketu P. Bhavsar}
\affil{Department of Physics and Astronomy, University of Kentucky, Lexington, KY 40506-0055, USA}
\email{bhavsar@pa.uky.edu}
\and
\author{Jatush V. Sheth}
\affil{Inter University Center for Astronomy and Astrophysics (IUCAA), Post Bag 4, Pune 411 007, India}
\email{jvs@iucaa.ernet.in}
\begin{abstract}
  
  We analyze the filamentarity in the Las Campanas redshift survey
  (LCRS) and determine the length scale at which filaments are
  statistically significant.  The largest length-scale at which
  filaments are statistically significant, real objects, is between
  $70$ to $80 \, h^{-1}$Mpc, for the LCRS $-3^o$ slice. Filamentary
  features longer than $80 \, h^{-1}$Mpc, though identified, are not
  statistically significant; they arise from chance alignments.  For
  the five other LCRS slices, filaments of lengths $50 \, h^{-1}$Mpc
  to $70 \, h^{-1}$Mpc are statistically significant, but not beyond.
  These results indicate that while individual filaments up to $80 \,
  h^{-1}$Mpc are statistically significant, the impression of
  structure on larger scales is a visual effect. On scales larger than
  $80 \, h^{-1}$Mpc the filaments interconnect by statistical chance
  to form the the filament-void network. The reality of the $80 \,
  h^{-1}$Mpc features in the $-3^o$ slice make them the longest
  coherent features in the LCRS.  While filaments are a natural
  outcome of gravitational instability, any numerical model attempting
  to describe the formation of large scale structure in the universe
  must produce coherent structures on scales that match these
  observations.

\end{abstract}

\newpage
\section{Introduction}

One of the most striking visual features in the distribution of
galaxies in the Las Campanas redshift Survey (LCRS, Shectman et al.
1996) is that they appear to be distributed along filaments.  These
filaments are interconnected and form a network, with voids largely
devoid of galaxies comprising the region in between the filaments.
This network of interconnected filaments encircling voids extends
across the entire survey and may be referred to as the ``Cosmic Web''.
Similar networks of filaments and voids are also visible in other
galaxy surveys, e.g., CfA (Geller \& Huchra 1989), 2dFGRS (Colless et
al. 2001, Colless et al. 2003) and SDSS (EDR) (Stoughton et al. 2002,
Abazajian et al. 2003). These, if they are a genuine feature of the
galaxy distribution, represent the largest structural elements in the
hierarchy of structures observed in the universe, namely galaxies,
clusters, superclusters, filaments and the cosmic web.

The analysis of filamentary patterns in the galaxy distribution has a
long history dating back to papers by Zel'dovich, Einasto and
Shandarin (1982), Shandarin and Zel'dovich (1983) and Einasto et al.
(1984). In the last paper the authors analyze the distribution of
galaxies in the Local Supercluster. They use the Friend-of-Friend
algorithm with varying neighborhood radius to identify connected
systems of galaxies referred to as ``clusters''.  As they increase the
neighborhood radius, they find that the clusters which are initially
spherical become multi-branched with multiple filaments of lengths up
to several tens of $h^{-1}$Mpc extending out in different directions.
Finally, as the radius is increased further, they find that the
filaments get interconnected and join neighboring superclusters into
an infinite network of superclusters and voids.  A later study
(Shandarin \& Yess 2000) used percolation analysis to arrive at a
similar conclusion for the distribution of the LCRS galaxies. The
large-scale and super large-scale structures in the distribution of
the LCRS galaxies have also been studied by Doroshkevich et al. (2001)
and Doroshkevich et al. (1996) who find evidence for a network of
sheet like structures which surround underdense regions (voids) and
are criss-crossed by filaments.  The distribution of voids in the LCRS
has been studied by M{\" u}ller, Arbabi-Bidgoli, Einasto \& Tucker
(2000) and the topology of the LCRS by Trac, Mitsouras, Hickson \&
Brandenberger (2002) and Colley (1997).  A recent analysis (Einasto et
al. 2003a) indicates a supercluster-void network in the Sloan Digital
Sky Survey also.

Traditionally, correlation functions (Peebles 1993) have been used to
quantify the statistical properties of the galaxy distribution. For
the LCRS, the two-point correlation function is a power law,
\b
\xi(r) = \left({r\over r_o}\right)^{-1.52}
\e
with the correlation length $r_o = 6.28$ $h^{-1}$Mpc on scales $2.0 \,
h^{-1}$Mpc to $16.4 \, h^{-1}$Mpc.  On scales larger than $30-40 \,
h^{-1}$Mpc, $\xi$(r) fluctuates closely around zero indicating a
statistically homogeneous galaxy distribution at and beyond these
scales (Tucker et al. 1997). This raises the question whether the
filamentary features which appear to span scales larger than $100 \,
h^{-1}$Mpc are statistically significant features of the galaxy
distribution or if they are mere artifacts arising from chance
alignment of the galaxies.

A quantitative estimator of filamentary structure, Shapefinder, was
defined (Bharadwaj et al. 2000) to provide a measure of the average
filamentarity for a point distribution in 2D. (See Sahni,
Sathyaprakash \& Shandarin 1998 for general introduction to
Shapefinders and Sheth et al. 2003 for the application of Shapefinders
to 3D simulations of structure formation.)  The Shapefinder statistic
was used to demonstrate that the galaxy distribution in the LCRS
exhibits a high degree of filamentarity compared to a random Poisson
distribution having the same geometry and selection effects as the
survey. This analysis provides objective confirmation of the visual
impression that the galaxies {\em are} distributed along filaments.
This, however, does not establish the statistical significance of the
filaments. The features identified as filaments are essentially chains
of galaxies, a crucial requirement being that the spacing between any
two successive galaxies along a chain is significantly smaller than
the mean inter-galaxy separation.  A chain runs as long as it is
possible to find another nearby galaxy which is not yet a member of
the chain, and breaks when no such galaxy is to be found. The fact
that the LCRS galaxies are highly clustered on small scales increases
the probability of finding pairs of galaxies at small separations.
This enhances the occurrence of long chains of galaxies, and we expect
to find a higher degree of filamentarity arising just from chance
alignments in the LCRS compared to a Poisson distribution.  To
establish whether the observed filaments are statistically significant
or if they are a result of chance alignments of smaller structural
elements, it is necessary to compare the sample of galaxies (here, the
LCRS slices) with a distribution of points which has the same small
scale clustering properties as the original sample and for which we
know that all large-scale filamentary features are solely due to
chance alignments. This is achieved using a statistical technique
called Shuffle (Bhavsar \& Ling 1988) whereby the statistical
significance of the filamentarity in a clustered dataset can be
assessed.

Shuffle generates fake data-sets, practically identical in their
clustering properties to the original data up to a length scale $L$,
but in which all structures longer than $L$, both real and chance, of
the original data, have been eliminated. In these Shuffled data,
filaments spanning length-scales larger than $L$ are visually evident,
even expected to be identified as a signal by the statistics used to
quantify the filamentarity, but all filaments spanning length-scales
larger than L have formed accidentally. The measure of the occurrence
of filaments spanning length-scales larger than $L$ in the Shuffled
data gives us a statistical estimate of the level at which chance
filaments spanning length-scales larger than $L$ occur in the original
data. Here we use Shuffle to estimate the degree of filamentarity
expected from chance alignments in the LCRS and use this to determine
the statistical significance of the observed filamentarity.

We present the method of analysis and our findings in Section 2.  In
Section 3 we discuss our results and present conclusions.

\section{Analysis and Results}
The LCRS contains the angular positions and redshifts of 26,418
galaxies distributed in 6 wedges, each $1.5^o$ thick in declination
and $80^o$ in right ascension. Three wedges are centered around mean
declinations $-3^o$, $-6^o$ and $-12^o$ in the Northern galactic cap
and three at declinations $-39^o$, $-42^o$ and $-45^o$ in the Southern
galactic cap. The survey has a magnitude limit $m = 17.75$ and extends
to a distance of $600 \, h^{-1}$Mpc. The most prominent visual feature
in these wedges (Figure 1) is that the galaxies appear to be
distributed along filaments, several of which span length-scales of
$100 \, h^{-1}$Mpc or more.

We extracted luminosity and volume limited sub-samples (Figure 1) from
the LCRS data so that we have an uniform sampling of the regions that
we analyze. In order to sample the largest regions that we could, with
the above criterion in mind, we limited the wedges from $195$ to $375
\, h^{-1}$Mpc in the radial direction as shown in Figure 1.

\begin{figure}
\figurenum{1}
\epsscale{0.5}
\rotatebox{-90}{\plotone{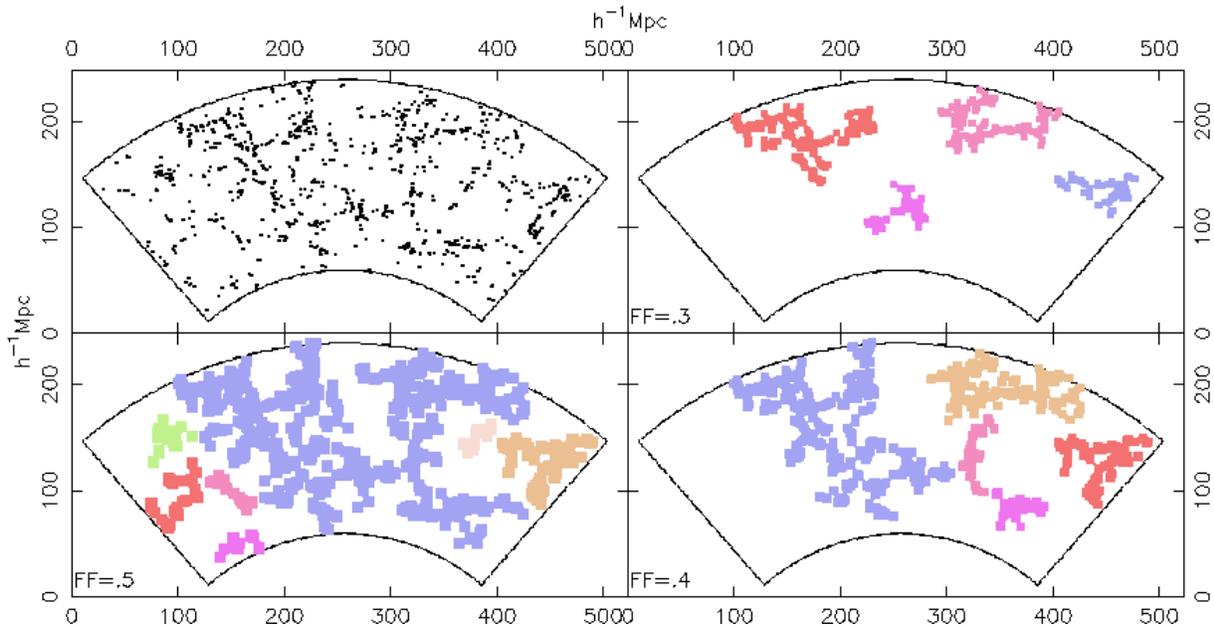}}
\caption{The upper left hand panel shows the distribution of galaxies
  in an uniformly sampled region of the $-3^o$ slice. The filamentary
  patterns of galaxies are evident. Going clockwise, the three other
  panels show some of the largest clusters identified using FOF after
  N=3,4 and 5 iterations of coarse-graining.  Most of the clusters
  shown are highly filamentary (${\cal F}_2 > 0.8$), with ${\cal F}_2
  > 0.9$ for the largest cluster in each panel (the definition of
  ${\cal F}_2$ is given in the text). At each level of
  coarse-graining, clusters grow until we have a single, large
  filamentary network spanning the whole region, referred to as the
  ``Cosmic Web''.
\label{fig1}}
\end{figure}

Our data consist of a total of 5073 galaxies distributed in 6 wedges.
Each LCRS wedge is collapsed along its thickness (in declination)
resulting in a 2 dimensional truncated conical slice which, being
geometrically flat, can be unrolled onto a plane. Each slice is
embedded in a $1 \, h^{-1}$Mpc $\times 1 \, h^{-1}$Mpc rectangular
grid.  Grid cells with galaxies in them are assigned the value 1,
empty cells 0. Connected regions of filled cells are identified as
clusters using a ``friends-of-friends''(FOF) algorithm. The geometry
and topology of each cluster is described by its area $S$, perimeter
$P$, and genus $G$. It is possible to utilize these measures to assess
the filamentarity of the supercluster of interest. This is achievable
using a set of measures termed as Shapefinders, originally defined for
2D hypersurfaces embedded in 3D. We use a 2D version of the
Shapefinder in our analysis of the superclusters in the LCRS slices.
However, before presenting our results, we digress briefly and
summarize the definition and the conceptual foundation of the
Shapefinder measures.

The geometry of a given structure is sensitive to any deformation
which the structure undergoes, while the topology of the structure,
solely relating to the connectedness of the structure, remains
unaffected. The morphology and the size of the objects is a result of
the interplay between both these aspects. The Shapefinders are
statistics devised to utilize both geometric and topological
information of a given object, to make a meaningful statement about
its size and its morphology. Both the geometry and topology of an
object are characterized in terms of Minkowski Functionals (hereafter
MFs). In 3D, the geometric MFs are (1) the Volume ${\cal V}$, (2) the
surface area ${\cal S}$ and (3) the integrated mean curvature ${\cal
  C}$, whereas the fourth MF is a topological invariant, the genus
${\cal G}$.  Sahni et al. (1998) defined three Shapefinders as the
ratio's of the above MFs, so as to have the dimensions of length.
These are conventionally considered to be reminiscent of the
characteristic thickness ${\cal T} =3{\cal V/S}$, breadth ${\cal B}
{\cal =S/C}$ and length ${\cal L} ={\cal C}/4\pi({\cal G}+1$) of the
object, and thus, together with genus ${\cal G}$, give us a feel for
the typical {\em size} and {\em topology} of the object of interest.
The information content about the three characteristic length-scales
associated with the object can further be used to make an objective
statement about the morphology of the object, as to how spherical,
planar, ribbon-like or filamentary an object is. Sahni et al. (1998)
proposed to achieve this by using two dimensionless Shapefinders,
namely planarity ${\cal P}$ and Filamentarity ${\cal F}$ defined as
follows:
\b
{\cal P} = {{\cal B-T}\over {\cal B+T}}; ~~~ {\cal F} = {{\cal L-B}\over {\cal L+B}}.
\e

These are defined so that for an ideal sheet like object, ${\cal P}$ =
1, ${\cal F}$ = 0, whereas for an ideal one-dimensional filament,
${\cal P}$ = 0, ${\cal F}$ = 1 {\footnote{The efficacy of the
    Shapefinder measures in revealing the morphology of the
    superclusters has been amply demonstrated (Sheth et al. 2003).
    These authors demonstrate that the percolating supercluster of the
    $\Lambda$CDM universe is the most filamentary, also consistent
    with its visual impression.  Further the more massive
    superclusters were shown to be more filamentary and the smaller
    structures were found to be quasi-spherical with ${\cal P}\sim
    {\cal F} \sim$0. These results prove the robustness of
    Shapefinders and confirm that the Shapefinders do indeed provide
    crucial insight into the morphology of the LSS, thus fulfilling
    the purpose for which these were devised.}}$^,${\footnote{One of
    the first morphological survey of the real Universe was done by
    Sathyaprakash et al. (1998) who analyzed the 1.2 Jy Redshift Survey
    Catalog.}}.

In our present analysis, we use the 2D version of the Shapefinders,
the Shapefinder measure ${\cal F}$
\b
{\cal F} = {({\cal P}^2 - 16{\cal S}) \over ({\cal P}-4)^2},
\e
originally defined in Bharadwaj et al. (2000), to quantify the shape
of the superclusters in the quasi-2D slices of the LCRS. By definition
0$\le {\cal F} \le$ 1.  ${\cal F}$ quantifies the degree of
filamentarity of a cluster, with ${\cal F}$ = 1 indicating a filament
and ${\cal F}$ = 0, a square (while dealing with a density field
defined on a grid).  The average filamentarity (${\cal F}_2$), is
defined as the mean filamentarity of all the clusters weighted by the
square of the area of the clusters.
\b {\cal F}_2 = {\sum_{i=1}^{N_{\rm cl}} {\cal S}_i^2 {\cal F}_i\over
  \sum_{i=1}^{N_{\rm cl}}{\cal S}_i^2} \e
In the current analysis, we use the average filamentarity to quantify
the degree of filamentarity in each of the LCRS slices.

The galaxy distribution in the LCRS slices is quite sparse and
therefore the Filling Factor $FF$ (defined as the fraction of filled
cells) is very small ($FF \sim 0.01$). The clusters identified using
FOF contain at most 2 or 3 filled cells, not yet corresponding to the
long filaments visually apparent in the slices. Larger structures are
identified by the method of ``coarse-graining''. Coarse-graining is
implemented by successively filling cells that are immediate neighbors
of already filled cells. It may be noted that the ``coarse-graining''
procedure adopted by us is equivalent to smoothing successively with a
top-hat kernel.  The filled cells get fatter after every iteration of
coarse-graining. This causes clusters to grow, first because of the
growth of filled cells, and then by the merger of adjacent clusters as
they overlap.  The observed large scale patterns are initially
enhanced as the clusters grow and then washed away as the clusters
become very thick and fill up the entire region.  $FF$ increases from
$FF \sim$ 0.01 to $FF =$ 1 as the coarse graining proceeds.  So as not
to limit ourselves to an arbitrarily chosen value of $FF$ as the one
defining filaments, we present our results showing the average
filamentarity for the entire range of filling factor $FF$.
\begin{figure}
\figurenum{2}
\epsscale{1.0}
\plotone{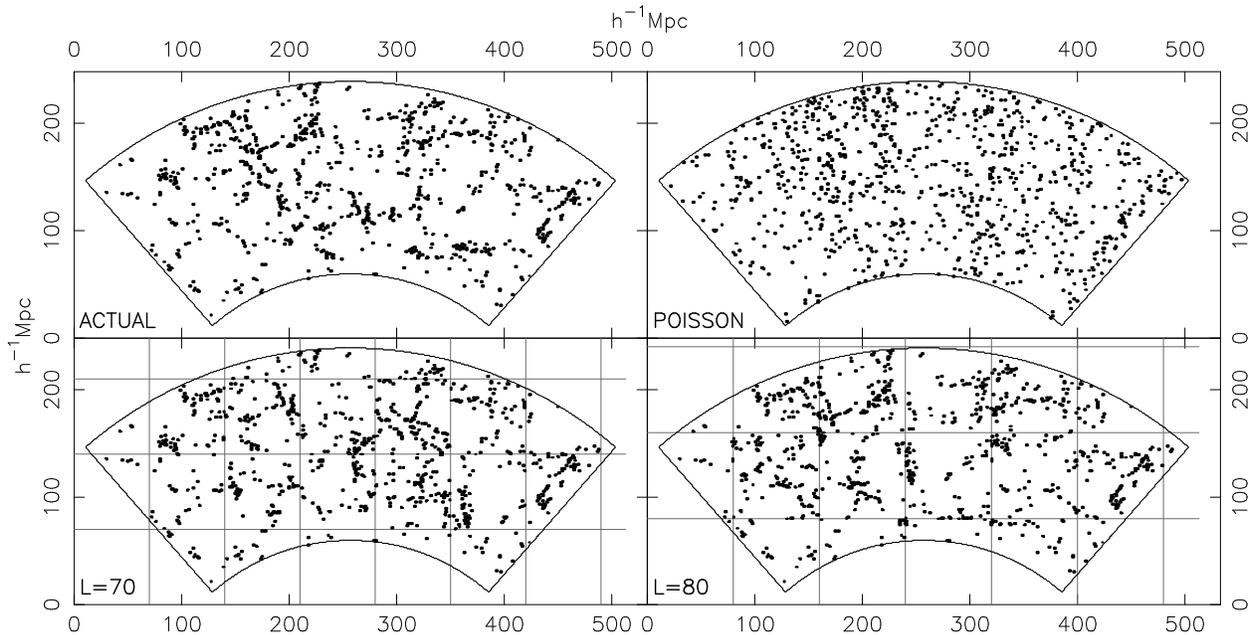}
\caption{The  data shown at top left, are the same as in Figure 1.
  The next panel (top right) shows a Poisson distribution of points
  generated over the same region. The bottom panels show Shuffled
  realizations of the data with $L = 70$ and $80 \, h^{-1}$Mpc. The
  square patches show the boundaries of the Shuffled regions While it
  is evident that the Poisson data are much less filamentarity than
  the LCRS galaxies, it is not possible to visually distinguish the
  level of filamentarity in the actual data from the Shuffled
  realizations.  A quantitative analysis, shows that the $L = 70 \,
  h^{-1}$Mpc Shuffled data exhibit less filamentarity than the actual
  data, while the $L = 80 \, h^{-1}$Mpc Shuffled data are
  statistically identical to the original data in filamentarity (fig.
  3). Note that all visual features across the boundaries in the
  bottom two panels (Shuffled data) are chance filaments.
\label{fig2}}
\end{figure}

Now we describe how the Shuffle algorithm works (Figure 2). A grid
with squares of side $L$ is superposed on the original data slice.
Square blocks of data which lie entirely within the slice are then
randomly interchanged, with rotation, repeatedly, to form a new
Shuffled slice. This process eliminates features in the original data
on scales longer than $L$, keeping clustering at scales below $L$
nearly identical to the original data. All the structures spanning
length-scales greater than $L$ that exist in the Shuffled slices are
the result of chance alignments. For each value of $L$ we use
different realizations of the Shuffled slices to estimate the degree
of filamentarity that arises from chance alignments on scales larger
than $L$. The Shuffled slices were analyzed in exactly the same way as
the actual LCRS slices. At a fixed value of $L$, the average
filamentarity in the {\em original} sample will be larger than in the
{\em Shuffled} data only if the actual data have more filaments
spanning length-scales larger than $L$, than that expected from chance
alignments. We vary the value of $L$ from $10 \, h^{-1}$Mpc to $100 \,
h^{-1}$Mpc and determine the largest value of $L$ ($L_{\rm MAX}$) such
that for all $L<L_{\rm MAX}$ the values of the average filamentarity,
${\cal F}_2$, in the actual data are higher than the Shuffled data,
indicating the presence of physical filaments of lengths greater than
$L$. We use three realizations for $L = 10,20,30,90$ and $100 \,
h^{-1}$Mpc shuffling of slices, and six realizations for the
intermediate length-scales. For length-scales beyond $L_{\rm MAX}$ the
average filamentarity in the Shuffled slices should continue to be the
same as in the actual LCRS slice, establishing $L_{\rm MAX}$ to be the
largest length-scale across which we have statistically significant
filamentarity.  Filaments which extend across length scales larger
than $L_{\rm MAX}$ are not statistically significant and are a
consequence of chance alignments. In Figure 3, we plot average
filamentarity, ${\cal F}_2$, as a function of the filling factor,
$FF$, for both the original sample as well as for samples generated by
shuffling the patches for various values of $L$. To convey the
essential aspects of the analysis, we only show the results for the
slices shuffled at our lowest $L$ value, $L = 10 \, h^{-1}$Mpc and at
or near the length-scale of interest, $L_{\rm MAX}$.
\begin{figure}
\figurenum{3}
\epsscale{0.8}
\rotatebox{-90}{\plotone{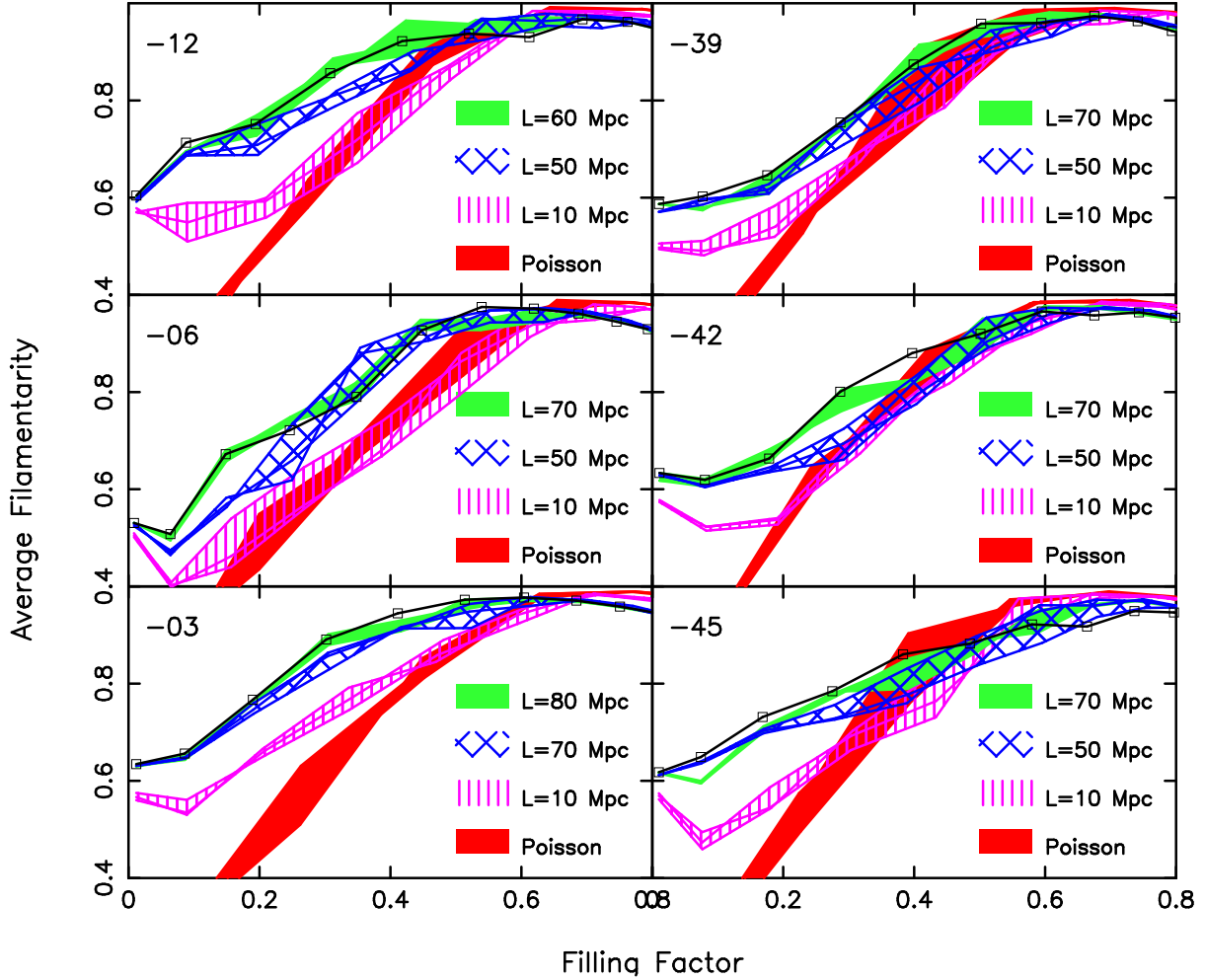}}
\caption{Plots of average filamentarity, ${\cal F}_2$, vs. Filling Factor, $FF$,
  for each of the LCRS slices (dark black line), three Shuffled
  realizations of each slice at various $L$ (hatched regions and light
  solid region)and three Poisson point distributions (dark solid
  region): We show Shuffle results for $L = 10 \, h^{-1}$Mpc, the
  smallest patches that were Shuffled, and the two L values where a
  transition occurs from the Shuffled galaxy distribution being less
  filamentary than the original data to it being indistinguishable
  from the original data. For each slice, this establishes $L_{\rm
    MAX}$, the length beyond which filaments are just chance
  artifacts, somewhere between these two values of $L$.
\label{fig3}}
\end{figure}
We use the $\chi^2$ test to establish $L_{\rm MAX}$ for the six
individual slices.  The reduced $\chi^2$ for the curves in Figure 3
are defined by
\b \chi^2(L) = {1\over
  N_B-1}\sum_{i=1}^{N_B}\l[{{\cal F}_2^{(i),{\rm LCRS}}-{\cal F}_2^{(i),{\rm Shuffled}}(L)\over\sigma_{{\cal F}_2}^{(i)}(L)}\r]^2,
\e
where $N_B$ is the number of data points available for comparison
between the original slice and a shuffled slice for a value of $L$.
The quantity $\sigma_{{\cal F}_2}^{(i)}(L)$ is the standard deviation
in ${\cal F}_2$ measured at a given filling factor, $FF_i$, using all
the available shuffled realizations at a given length scale $L$. We
have noted that for filling factor, $FF>0.7$, all the ${\cal F}_2$
curves follow the same trend, regardless of the slice (original or
shuffled).  This can be interpreted as the regime of $FF$ in which the
coarse-graining defines structures of such large extent that they are
unphysical and Shuffling does not discriminate between original or
Shuffled data. Including the tiny differences in the curves in this
regime ($FF>0.7$), will give weight to an unphysical signal and
determine an erroneous $\chi^2$.  Hence, we only include the region of
the curves for $FF<0.7$ to determine the reduced $\chi^2$, using this
as the discriminating measure between the curve for the real data and
the Shuffled realizations at different $L$. The reduced $\chi^2$
quantifies how different a Shuffled slice is from the original, at
various $L$. The minimum value of the reduced $\chi^2$ should
correspond to the length-scale $L=L_{\rm MAX}$ at which, if slices are
Shuffled, the filamentarity of the Shuffled slices and the original
slice differ the least. This gives us the length scale $L_{\rm MAX}$
beyond which filaments are only chance objects and not physical.
\begin{figure}
\figurenum{4}
\epsscale{0.8}
\rotatebox{0}{\plotone{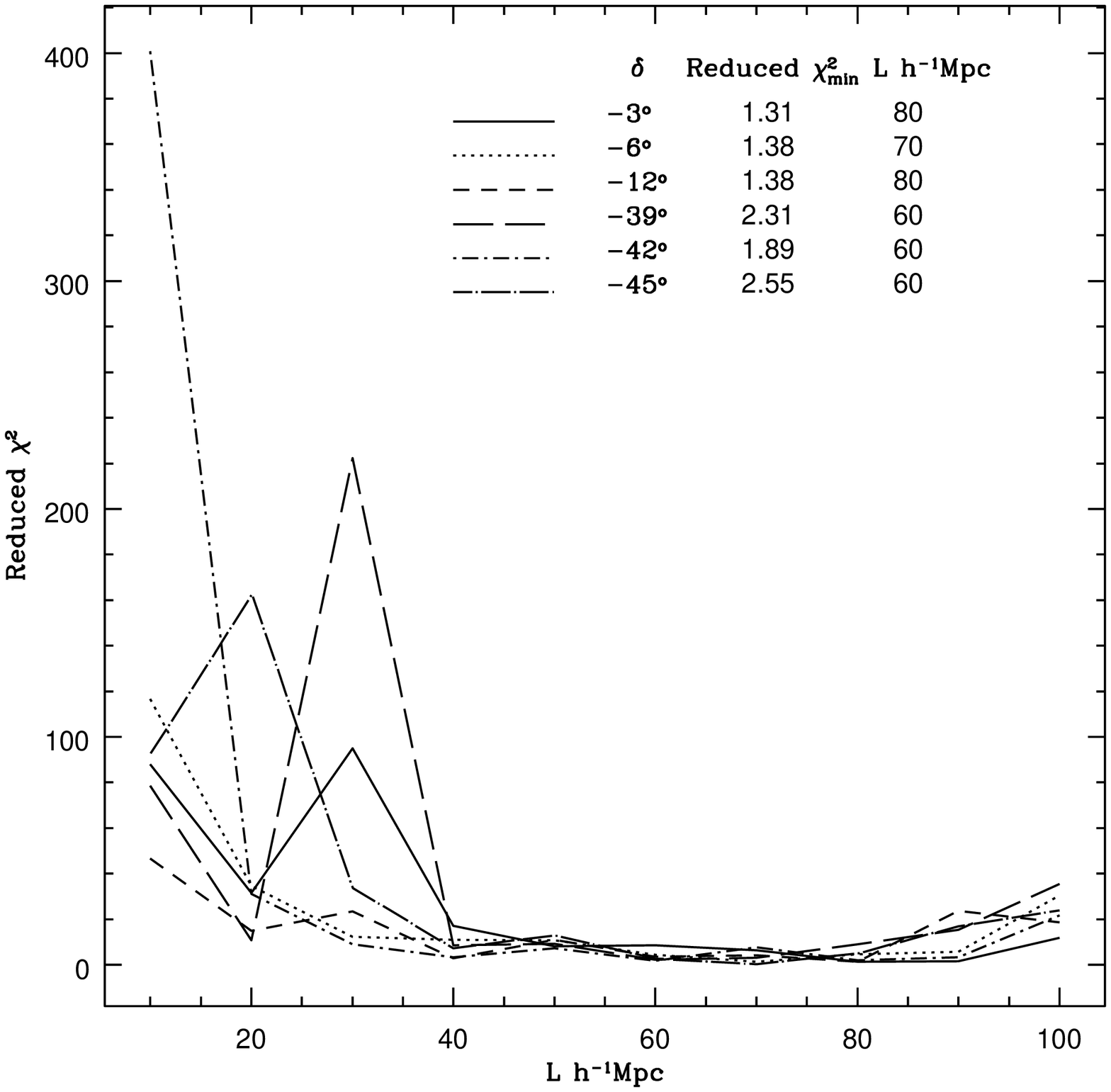}}
\caption{The minimum value of the reduced $\chi^2$($L$) is plotted for all the
  six slices. It varies from 1.4 to 2.6 and is in within the
  acceptable bounds. We conclude from here that the scale of longest
  {\em real} filaments is $\sim 60 \, h^{-1}$Mpc for all the three
  southern slices. For the northern slices this scale is $\sim 70-80
  \, h^{-1}$Mpc. This scale can also be interpreted as the scale
  beyond which the LSS in the Universe is homogeneous.
  \label{fig4}}
\end{figure}

In Figure 4, we show the reduced $\chi^2$ vs. $L$ plotted for the six
slices.  We also list the minimum values of reduced $\chi^2$ for each
slice and the corresponding $L$. We see that the values of reduced
$\chi^2$ are well within acceptable bounds to say that the Shuffled
slices at these values of $L$ are indistinguishable from the original
slice. The length-scale that correspond to these minima is $\sim 60 \,
h^{-1}$Mpc for all the southern slices, whereas it is $\sim 70 \,
h^{-1}$Mpc for $-6^o$ slice and $80 \, h^{-1}$Mpc for $-3^o$ and
$-12^o$ slices.  We thus establish that for the Southern slices the
longest real filaments are no longer than $60 \, h^{-1}$Mpc and for
the Northern slices no longer than $80 \, h^{-1}$Mpc. Beyond $80 \,
h^{-1}$Mpc structure is not statistically significant.

\section{Discussion and conclusions}
A look at the filamentary features at different levels of
coarse-graining (Figure 1) reveals that the size of the largest
filamentary feature increases monotonically with successive iterations
of coarse-graining until it spans the entire survey (Bharadwaj et al.
2000). As coarse-graining proceeds, individual filaments form and then
interconnect to form the supercluster-void network, in keeping with
the earlier analysis (Einasto et al. 1984; Shandarin \& Yess 2000;
Einasto et al. 2003a) discussed in the introduction.  Although, the
length of the interconnected network of filaments increases
monotonically, the ratio of the length to the number of holes (Genus)
stabilizes and then decreases (Bharadwaj et al. 2000). For the
$-3^{o}$ slice, this ratio stabilizes around $140 \, h^{-1}$Mpc at $FF
\sim$ 0.4. This ratio may be interpreted as the perimeter of the
typical void in the network. This leads to a picture where there are
voids of diameter $60 \, h^{-1}$Mpc encircled by filaments of thickness
$10 \, h^{-1}$Mpc (Peebles 1993, Sheth et al. 2003, Sheth 2003)
interconnected to form a large web. A void of this size, along with
the filament at its perimeter, would span a length-scale $\sim 80 \,
h^{-1}$Mpc. The results of this paper show that such voids encircled
by filaments are statistically significant features. Although our
analysis also finds a web of interconnected filaments which spans
length-scales larger than $80 \, h^{-1}$Mpc and runs across the entire
survey, this is not statistically significant. The web arises from
chance interconnections between the filaments encircling different
voids.

Studies of the distribution of Abell superclusters (Einasto et al.
1997b) show that the mean distance between neighboring superclusters
is about $50 \, h^{-1}$Mpc for poor superclusters and about $100 \,
h^{-1}$Mpc for rich superclusters. The distribution of the SDSS
superclusters (Einasto et al. 2003a) and the LCRS superclusters
(Einasto et al.  2003b) shows a similar behavior.  Visualizing the
superclusters as being randomly distributed, we would expect filaments
joining the superclusters to develop as the density field is
progressively smoothened.  Such filaments will arise from the chance
alignments of shorter, genuine, statistically significant filaments.
The filaments joining superclusters will span length-scales comparable
to the mean inter-supercluster separation and the statistical
properties of filaments would be stable to shuffling {\it i.e.}, it
would not change if the superclusters were rearranged randomly. Our
results may be interpreted as being indicative of the superclusters
being randomly distributed on scales larger than $80 \, h^{-1}$Mpc
with the mean inter supercluster separation also being of this order
{\footnote{ It is interesting to note that a study of the SDSS(EDR)
    superclusters conducted by Doroshkevich et al. (2003) using
    Minimal Spanning Trees concludes that the large-scale filaments
    appear to randomly connect the sheet-like structures in more
    denser environments.}}.

The presence of statistically significant features on scales as large
as $70$ to $80 \, h^{-1}$Mpc may seem surprising given the fact that
the correlation analysis fails to detect any clustering on scales
beyond $30$ to $40 \, h^{-1}$Mpc. This is due to the inability of the
two point (and higher order) correlation functions in detecting
coherence at large-scales.  Pattern specific methods (like
Shapefinders) are necessary to detect and quantify coherent large
scale features in the galaxy distribution. It is interesting to note
that the two dimensional power spectrum for the LCRS (Landy et al.
1996) exhibits strong excess power at wavelengths $\sim 100 \,
h^{-1}$Mpc, a feature which may possibly be related to the filamentary
patterns studied here.  The analysis of the three dimensional
distribution of Abell clusters (Einasto et al. 1997a, 1997b) reveals a
bump at $k=0.05 \, h \, {\rm Mpc}^{-1}$ in the power spectrum.  Also,
the recent analysis of the SDSS shows a bump at $k=0.05 \, h \, {\rm
  Mpc}^{-1}$ in the power spectrum (Tegmark et al.  2003). While it is
interesting to conjecture that these features may be related to the
presence of filaments, we should also note that the filaments are
non-Gaussian features and cannot be characterized by the power
spectrum alone.

A point which should be noted is that the filaments quite often run in
a zig-zag fashion (Figure 1), and the length of a filament which spans
a length-scale of $80 \, h^{-1}$Mpc may be significantly larger than
$80 \, h^{-1}$Mpc. Also, the present analysis is two dimensional
whereas the filaments actually extend in all three dimensions. The
length of the filaments may be somewhat larger in three dimensions,
and a little bit of caution may be advocated in generalizing our
results {\footnote{In this context, it is interesting to note that in
    a recent analysis of mock SDSS catalogs based on $\Lambda$CDM
    model, Sheth (2003) finds the length-scales of the largest
    superclusters to be $\sim 60 \, h^{-1}$Mpc. This indicates as to
    what might be anticipated in extending this work to 3D redshift
    surveys.}}.

In the gravitational instability picture, small disturbances in an
initially uniform matter distribution grow to produce the large-scale
structures presently observed in the universe. It is possible to
interpret the filaments in terms of the coherence of the deformation
(or strain) tensor (Bond, Kofman \& Pogosyan 1996) of the smoothened
map from the initial to the present positions of the particles which
constitutes the matter. Our analysis shows that the deformation tensor
has correlation to length-scales up to $80 \, h^{-1}$Mpc and is
uncorrelated on scales larger than this. The ability to produce
statistically significant filamentarity on scales up to $80 \,
h^{-1}$Mpc will be a crucial quantitative test of the different models
for the formation of Large Scale Structure in the universe.

We next address the question of the length-scale beyond which the
distribution of galaxies in the LCRS may be considered to be
homogeneous. The analysis of Kurokawa, Morikawa \& Mouri (2001) shows
this to occur at a length-scale of $\sim 30 \, h^{-1}\,{\rm Mpc}$,
whereas Best (2000) fails to find a transition to homogeneity even on
the largest scale analyzed.  The analysis of Amendola \& Palladino
(1999) shows a fractal behavior on scales less than $\sim$
$30 h^{-1}\,{\rm Mpc}$ but is inconclusive about the transition to
homogeneity. The results presented in this paper set a lower limit to
this length-scale at around $80 \, h^{-1}$Mpc, in keeping with estimates
based on the multi-fractal analysis of LCRS (Bharadwaj, Gupta \&
Seshadri 1999) who find that the LCRS exhibits homogeneity on the
scales $80$ to $200 \, h^{-1}\, {\rm Mpc}$. In a separate approach
based on the analysis of the two point correlation applied to actual
data and simulations Einasto and Gramann (1993) find that the
transition to homogeneity occurs at about $130 \, h^{-1} {\rm Mpc}$.  For
the LCRS, the scale of the largest coherent structure is at least
twice the length-scale at which the two-point correlation function
becomes zero. Beyond this scale the filaments interconnect
statistically to form a percolating network.  This filament-void
network of galaxies is not distinguishable, in a statistical sense,
beyond scales of $80 \, h^{-1}$Mpc. If the LCRS slices can be considered
a fair sample of the universe then this suggests the scale of
homogeneity for the universe.

\section{Acknowledgment}
The authors wish to thank the LCRS team for making the survey data
public. SB would like to acknowledge financial support from the Govt.
of India, Department of Science and Technology (SP/S2/K-05/2001). SPB
would like to thank the Kentucky Space Grant Consortium (KSGC)for
funding. JVS is supported by the senior research fellowship of the
Council of Scientific and Industrial Research (CSIR), India. We also
wish to thank the referee Jaan Einasto for providing very valuable and
useful comments on our manuscript.

\end{document}